\documentclass[twocolumn]{aastex63}
\usepackage{bm}
\usepackage{soul}
\usepackage{xfrac}
\usepackage{xcolor}
\usepackage{amsmath}
\usepackage{CJKutf8}
\usepackage{hyperref}
\usepackage{textcomp}
\usepackage{multirow}
\usepackage{graphicx}
\usepackage{booktabs}
\usepackage{tabularx}
\usepackage{rotating}
\usepackage{makecell}
\usepackage{mathrsfs}
\usepackage{ragged2e}
\usepackage{mathtools}
\usepackage[T1]{fontenc}
\usepackage[utf8]{inputenc}
\usepackage[normalem]{ulem}
\usepackage[makeroom]{cancel}

\newsavebox{\savefig}

\newcommand{\V}{\textit{\textbf{V}}}
\newcommand{\W}{\textit{\textbf{W}}}

\newcommand{\simpchinese}[1]{\begin{CJK}{UTF8}{gbsn}#1\end{CJK}}
\newcommand{\tradchinese}[1]{\begin{CJK}{UTF8}{bsmi}#1\end{CJK}}

\newcolumntype{C}[1]{>{\centering\arraybackslash}m{#1\dimexpr0.024\textwidth}} % for full-width table

\begin{document}

\title{Shoulder of Dust Rings Formed by Planet-disk Interactions}

\author[0000-0002-0605-4961]{Jiaqing Bi \simpchinese{(毕嘉擎)}}
\affil{Academia Sinica Institute of Astronomy \& Astrophysics, Taipei 106216, Taiwan, Republic of China}
\affil{Department of Astronomy \& Astrophysics, University of Toronto, Toronto, ON M5S 3H4, Canada}
\email{jbi@asiaa.sinica.edu.tw}

\author[0000-0002-8597-4386]{Min-Kai Lin \tradchinese{(林明楷)}}
\affil{Academia Sinica Institute of Astronomy \& Astrophysics, Taipei 106216, Taiwan, Republic of China}
\affil{Physics Division, National Center for Theoretical Sciences, Taipei 106216, Taiwan, Republic of China}
\email{mklin@asiaa.sinica.edu.tw}

%%%%%%%%%%%%%%%%%%%%%%%%%%%%%%%%%%%%%%%%%%%%%%%%%%%%%%%%%%%%%%%%%%%%%%%%%%%%%%%%
%%%%%%%%%%%%%%%%%%%%%%%%%%%%%%%%%%%%%%%%%%%%%%%%%%%%%%%%%%%%%%%%%%%%%%%%%%%%%%%%

\begin{abstract}
    
Recent analyses of mm-wavelength protoplanetary disk observations have revealed several emission excesses on the previously identified dust rings, referred to as dust shoulders. The prevalence of dust shoulders suggests that they trace a common but unclear mechanism. In this work, we combine 3D, multifluid hydrodynamic simulations with radiative transfer calculations to explain the formation of dust shoulders. We find that the ring-shoulder pairs can result from the 3D planet-disk interactions with massive, gap-opening planets. The key driver is the dust filtration effect at the local pressure maximum due to planet-driven outward gas flows. Our work provides a possible explanation for the outer dust shoulders in recent super-resolution analyses of ALMA observations. It also provides insights into the formation of the inner dust shoulder in the PDS 70 disk and highlights the role of 3D effects in planet-disk interaction studies.

\end{abstract}


%%%%%%%%%%%%%%%%%%%%%%%%%%%%%%%%%%%%%%%%%%%%%%%%%%%%%%%%%%%%%%%%%%%%%%%%%%%%%%%%
%%%%%%%%%%%%%%%%%%%%%%%%%%%%%%%%%%%%%%%%%%%%%%%%%%%%%%%%%%%%%%%%%%%%%%%%%%%%%%%%

\section{Introduction} \label{sec:intro}

Our current understanding of planet formation theory suggests that dust components in protoplanetary disks play a critical role in the planet-forming process (see \citealt{birnstiel_dust_2023} for a review). In the meantime, as one of the primary sources of emission in the disk, dust provides abundant information on disk dynamics via the nontrivial, local enhancement and depletion (substructures) observed in millimeter wavelengths. Among those dust substructures, surveys performed by the Atacama Large Millimeter/submillimeter Array (ALMA) revealed that annular rings and gaps are the most prevalent kind (e.g., \citealt{huang_disk_2018-1, long_gaps_2018}). 

The prevalence of rings and gaps is exciting news because they may be correlated to the formation of planets, causally or consequentially. On the one hand, planets form rings and gaps because massive planets tend to open a gap at their orbital region via gravitational interactions with the disk (e.g., \citealt{lin_tidal_1993}). Then, local pressure maxima formed at the gap edges can trap dust grains into rings via aerodynamic effects (e.g., \citealt{takeuchi_dust_2001}). On the other hand, dust rings form planets because the high concentration of dust there promotes dust growth via a series of mechanisms such as collisional growth, streaming instability, and pebble accretion (see \citealt{drazkowska_planet_2023} for a review).

Planet-disk interactions are not the only explanation for dust rings and gaps. Several studies have discovered other mechanisms such as zonal flows from magneto-hydrodynamic turbulence (e.g., \citealt{johansen_zonal_2009}), chemistry-related evolution of dust (e.g., \citealt{zhang_evidence_2015, okuzumi_sintering-induced_2016}), various dynamical instabilities (e.g., \citealt{takahashi_two-component_2014, dullemond_dust-driven_2018}), and dust feedback onto the gas (e.g., \citealt{jiang_survival_2021}). A pathway to distinguish planet-origin rings and gaps from others is critical to benefit the current planet-hunting campaign.

In \citealt{bi_puffed-up_2021} and \citealt{bi_gap-opening_2023}, we sought characteristic features of dust substructures under planet-disk interactions using three-dimensional (3D), global hydrodynamic simulations with both gas and dust species. We found the dust layer of sub-mm-sized grains, which is conventionally assumed to settle on the disk midplane, puffs up at the edges of the planet-opened gap. The puff-up results from the encounter between the gap-opening flow driven by the planet and the gap-closing flow driven by the viscous evolution at the gap edges (\citealt{fung_gap_2016}). The two flows merge, head upwards, and carry dust grains to high disk elevations. Moreover, we found that the sub-mm-sized grains, considered moderately coupled to the gas, are susceptible to an effective diffusion resulting from the long-term periodic perturbations from the planet-driven spiral waves. As a result, the dust rings under planet-disk interactions tend to be more radially extended than those with other origins in mm-wavelength observations. 

The change in vertical and radial distributions of dust resulting from planet-disk interactions is a distinctive indicator of planets in the disk. However, it is essential to note that these phenomena rely on the gas flows driven by the planet, thus requiring a certain level of coupling between gas and dust. In other words, they are dependent on the grain size. Therefore, in this follow-up study, we focus on the spatial distribution of dust when multiple dust species are considered simultaneously. Furthermore, we conduct radiative transfer analyses to offer insights into mm-wavelength disk observations. 

Meanwhile, thanks to the high-resolution observations by ALMA and the development of data processing techniques, this work is motivated by the discovery of emission excesses (a.k.a., shoulders) on several previously identified dust rings. For example, \citealt{huang_multifrequency_2020} identified a shoulder on the inner dust ring of the GM Aur disk at 2.1 mm wavelength using a 57$\times$34 mas beam (9$\times$5 au). In the PDS 70 disk, a shoulder was identified by \citealt{keppler_highly_2019} in the semi-major axis profile at 0.9 mm, which was confirmed and further suggested to be an axisymmetric feature by \citealt{benisty_circumplanetary_2021} using higher resolution observations and the tool \textsc{frank} (\citealt{jennings_frankenstein_2020}). Then, \citealt{jennings_super-resolution_2022} and \citealt{jennings_superresolution_2022} revealed even more shoulders in several other disks using super-resolution analyses of the archival data, suggesting a common but obscured origin of shoulders lying behind. 

This work demonstrates that the shoulders in mm-wavelength dust observations may result from planet-disk interactions. The paper is organized as follows: In section \ref{sec:method}, we introduce the numerical methods involved in this work, including hydrodynamic simulations and radiative transfer calculations. In Section \ref{sec:result}, we demonstrate the results, which include the density and dust emission profiles resulting from the 3D planet-disk interactions. In Section \ref{sec:discuss}, we discuss the connections between our work and the shoulders found in previous observations. Finally, we conclude in Section \ref{sec:conclusion}.
\section{Numerical Methods} \label{sec:method}

The workflow is described as follows: We first run a gas-only hydrodynamic simulation with a planet until a gas gap is opened in the disk. We then add a dust ring at the local pressure maximum outside the gap and let the gas-and-dust simulation run further. Finally, we perform radiative transfer calculations for the hydrodynamic outputs. The hydrodynamic simulations and radiative transfer calculations are described in the following sections.

\subsection{Hydrodynamic Simulations}

We consider a 3D protoplanetary disk model with an embedded planet of mass $M_{\rm p}$ around a central star of mass $M_\star$. The planet is treated as a point mass with a softened potential, and the planet-disk model neglects disk self-gravity, magnetic fields, planet migration, and planet accretion. We use $\{r,\, \phi,\, \theta\}$ to denote spherical radius, azimuth, and polar angle, and $\{R,\, \phi,\, Z\}$ to denote cylindrical radius, azimuth, and height. Both coordinates are centered on the star. The subscript ``ref'' denotes evaluations on the midplane at the reference radius ($R = R_{\rm ref},\, Z = 0$). The subscript ``0'' denotes initial values.

\subsubsection{Basic Equations and Model Setups}

The basic equations for hydrodynamic models are the same as those in \citealt{bi_puffed-up_2021}. Therefore, we only provide a summary here and refer readers to the previous work for more detailed descriptions and justifications. 

The volumetric density, pressure, and velocity vector of gas are denoted by $\{\rho_{\rm g},\, P,\, \textit{\textbf{V}}_{\rm g}\}$. The isothermal equation of state is given by $P = \rho_{\rm g}c_{\rm s}^2$, where $c_{\rm s}(R) = c_{\rm s, ref}(R/R_{\rm ref})^{-1/2}$ is the sound speed of gas. We assume a time-independent, vertically isothermal, axisymmetric power-law gas temperature profile $\propto R^{-1}$. It gives a non-flared gas disk with a constant aspect ratio, for which we adopt a value of 0.05. We consider a constant gas kinematic viscosity $\nu = 10^{-5}R_{\rm ref}^2\Omega_{\rm K,ref}$, where $\Omega_{\rm K}(R) = (GM_\star/R)^{-1/2}$ is the Keplerian angular velocity and $G$ is the gravitational constant. It corresponds to $\alpha = 4\times10^{-3}$ at $R = R_{\rm ref}$ in the $\alpha$-viscosity prescription (\citealt{shakura_black_1973}). 

The volumetric density and velocity vector of dust are denoted by $\{\rho_{\rm d},\, \textit{\textbf{V}}_{\rm d}\}$. The dust components are modeled as pressureless fluids in our hydrodynamic simulations. Dust feedback onto the gas is considered, but dust-dust interactions (e.g., coagulation and fragmentation) and dust diffusion are neglected. The dust-gas coupling is parameterized by the Stokes number ${\rm St} = \tau_{\rm s}\Omega_{\rm K}$. We assume dust grains are in the Epstein regime with fixed grain sizes. The particle stopping time $\tau_{\rm s}$ is given by $$\tau_{\rm s} = \frac{\rho_{\rm g0, ref}}{\rho_{\rm g}}\frac{c_{\rm s, ref}}{c_{\rm s}}\frac{{\rm St}_{\rm ref}}{\Omega_{\rm K, ref}},$$ where ${\rm St}_{\rm ref}$ is a fixed parameter related to the grain size\footnote{We note that $\rho_{\rm g0,ref}$ is the unperturbed gas density at the reference radius before the planet opening a gap.}. Our hydrodynamic models simultaneously consider two species of dust with ${\rm St}_{\rm ref} = 10^{-3}$ and $10^{-1}$ (hereafter, the small dust grains and the large dust grains).

The hydrodynamic equations for gas and dust are given by 
\begin{align} 
    \label{eq:gascont}
    &\frac{\partial \rho_{\rm g}}{\partial t} + \nabla \cdot (\rho_{\rm g} \textit{\textbf{V}}_{\rm g}) = 0, \\
    &\begin{aligned}
        \frac{\partial \textit{\textbf{V}}_{\rm g}}{\partial t} + \textit{\textbf{V}}_{\rm g} \cdot \nabla \textit{\textbf{V}}_{\rm g} = &- \frac{1}{\rho_{\rm g}} \nabla P - \nabla \Phi \\
        &+ \frac{\epsilon}{\tau_{\rm s}}(\textit{\textbf{V}}_{\rm d} - \textit{\textbf{V}}_{\rm g}) + \frac{1}{\rho_{\rm g}} \nabla \cdot \mathcal{T},
    \end{aligned} \\
    \label{eq:dustcont}
    &\frac{\partial \rho_{\rm d}}{\partial t} + \nabla \cdot (\rho_{\rm d} \textit{\textbf{V}}_{\rm d}) = 0, \\ 
    &\frac{\partial \textit{\textbf{V}}_{\rm d}}{\partial t} + \textit{\textbf{V}}_{\rm d} \cdot \nabla \textit{\textbf{V}}_{\rm d} =  - \nabla \Phi - \frac{1}{\tau_{\rm s}}(\textit{\textbf{V}}_{\rm d} - \textit{\textbf{V}}_{\rm g}).
\end{align}
Here, $\Phi$ is the net potential, including terms from the star, the planet, and the indirect planet-star gravitational interactions (\citealt{regaly_circumstellar_2017}). $\epsilon$ is the volumetric dust-to-gas density ratio. And $\mathcal{T}$ is the viscous stress tensor related characterized by a constant kinematic viscosity $\nu$.

The planet is on a fixed circular orbit on the disk midplane at $R = R_{\rm ref}$. We consider two models with a Saturn-mass planet ($M_{\rm p} = 3\times10^{-4}M_\star$) in one and a Jupiter-mass planet ($M_{\rm p} = 1 \times 10^{-3}M_\star$) in the other. A smoothing length of 0.1$H_{\rm g}$ is used when calculating the planet's potential, where $H_{\rm g}$ is the gas scale height. 

\subsubsection{Initialization} \label{sec:dust_init}

The simulation starts with a gas-only disk. The axisymmetric gas density is initialized to 
\begin{equation}
    \rho_{\rm g0} = \rho_{\rm g0,ref} \left(\frac{R}{R_{\rm ref}}\right)^{-3/2} \times \exp\left[\frac{GM_\star}{c_{\rm s}^2}\left(\frac{1}{r} - \frac{1}{R}\right)\right],
\end{equation} 
where $\rho_{\rm g0,ref}$ can be arbitrary for a non-self-gravitating disk. The azimuthal velocity is initialized to $V_{\rm g,\phi 0} = (1-2\eta)^{1/2}R\Omega_{\rm K}$, where $\eta$ is a dimensionless measurement of the radial pressure gradient. The radial and vertical velocities are initialized to zero. The gas-only simulation lasts 500 $P_{\rm ref}$, where $P_{\rm ref}=2\pi\Omega_{\rm K,ref}^{-1}$ is the orbital period at the reference radius. It is sufficiently long for a planet-opened gas gap to reach steady width and depth. The planet's mass gradually increases from zero to $M_{\rm p}$ in the first 100 $P_{\rm ref}$. At the end of the gas-only phase, a local pressure maximum is formed at the outer gap edge.

A dust ring composed of two dust species is added to the disk at the local pressure maximum at t = 500 $P_{\rm ref}$. The dust ring is initialized via the volumetric dust-to-gas density ratio such that 
\begin{equation}
    \epsilon(R,Z,{\rm St_{ref}}) = \epsilon_0({\rm St_{ref}})\times\exp{\left[-\frac{(R-R_{\rm Pmax})^2}{2\sigma_R^2}\right]}\times\exp{\left(-\frac{Z^2}{2\sigma_Z^2}\right)},
\end{equation} 
where $R_{\rm Pmax}$ is the radius of the local pressure maximum, $\sigma_R = H_{\rm g}$, and $\sigma_Z = 0.2H_{\rm g}$. To investigate the dust feedback from different dust species, we explore three combinations of $\epsilon_0({\rm St_{ref} = 10^{-3}})$ and $\epsilon_0({\rm St_{ref} = 10^{-1}})$. They are (1) $\epsilon_0({\rm St_{ref} = 10^{-3}}) = 0.1$ and $\epsilon_0({\rm St_{ref} = 10^{-1}}) = 1$; (2) $\epsilon_0({\rm St_{ref} = 10^{-3}}) = \epsilon_0({\rm St_{ref} = 10^{-1}}) = 1$; and (3) $\epsilon_0({\rm St_{ref} = 10^{-3}}) = 1$ and $\epsilon_0({\rm St_{ref} = 10^{-1}}) = 0.1$. Therefore, dust feedback on the gas is expected to be non-negligible in all three models. The azimuthal velocity is initialized to $V_{\rm d,\phi 0} = (GM_\star/r)^{-1/2}$. The radial and vertical velocities are initialized to zero. The gas-and-dust phase runs for 3000 $P_{\rm ref}$ for the Saturn-mass planet model and 7000 $P_{\rm ref}$ for the Jupiter-mass planet model. We note that while 3000 $P_{\rm ref}$ is sufficient for dust substructures in the Saturn-mass planet model to reach a steady state, 7000 $P_{\rm ref}$ is not sufficient for the small dust in the Jupiter-mass planet model. Nevertheless, we show in Appendix \ref{app:time} that the non-steady state has limited effects on the dust emission profile and thus does not impact our interpretations and conclusions.

\subsubsection{Numerical Setup}

The model is evolved by \textsc{fargo3d} (\citealt{benitez-llambay_fargo3d_2016, benitez-llambay_asymptotically_2019}). We adopt a spherical domain centered on the central star from 0.4 to 4.0$R_{\rm ref}$ in $r$, full $2\pi$ in $\phi$, and $\pm3 H_{\rm g}$ equivalent in $\theta$. The grids are logarithmically spaced in $r$ and uniformly in $\phi$ and $\theta$. The grid resolutions are $N_{\theta}\times N_{r}\times N_{\phi} = 90\times450\times630$, which correspond to 15, 10, and 5 grids per $H_{\rm g}$, respectively.

Periodic conditions are applied at $\phi$ boundaries. Dust densities are symmetric at $r$ and $\theta$ boundaries. Gas density and all azimuthal velocities are extrapolated at $r$ and $\theta$ boundaries. All velocities in $\theta$ are anti-symmetric at $\theta$ boundaries and symmetric at $r$ boundaries. Except that the inner radial boundary is open for mass loss of dust, all velocities in $r$ are anti-symmetric at $r$ boundaries and symmetric at $\theta$ boundaries. Wave-killing conditions are applied to the gas at $r$ boundaries within a margin of 10\% of the boundary radius. 

\subsection{Radiative Transfer Calculations} \label{sec:radmc3d}

We consider a 3D, face-on protoplanetary disk at 140 pc surrounding a T Tauri star of mass $M_\star = 0.76 M_\odot$ and black body temperature $T_\star = 3972{\rm K}$ (i.e., same as those for PDS 70; \citealt{pecaut_star_2016, muller_orbital_2018, keppler_discovery_2018}). The radiative transfer model considers three dust species: dust grains of 1 $\mu$m, 100 $\mu$m, and 1 cm in size. They inherit the density distribution of gas\footnote{Here, we assume that the 1-$\mu$m-sized dust grains are well mixed with gas (i.e., perfectly coupled) in the disk.}, ${\rm St}_{\rm ref} = 10^{-3}$ dust, and ${\rm St}_{\rm ref} = 10^{-1}$ dust in our hydrodynamic simulations, respectively. The grain size and Stokes number are bridged by the assumption of dust grains with an internal density of $\sim$1 g/cm$^{3}$ at $\sim$50 au in an MMSN-like disk (\citealt{hayashi_structure_1981}). 

We use \textsc{radmc3d} (\citealt{dullemond_radmc-3d_2012}) to calculate the dust continuum emission at 1.3 mm wavelength. The numerical grids are identical to those in hydrodynamic simulations and are scaled such that the planet is at 50 au. We use $10^{10}$ photons for both thermal and scattering Monte Carlo, sufficient to provide convergence in 3D calculations. We adopt anisotropic scattering using the Henyey-Greenstein function (\citealt{henyey_diffuse_1941}). The thermal Monte Carlo is performed in the wavelength range between 0.1 $\mu$m and 1 cm with 200 points. Dust opacity at each wavelength point is derived using \textsc{optool} (\citealt{dominik_optool_2021}) and the grain composition from the DSHARP survey (\citealt{birnstiel_disk_2018}). Finally, the 2D dust emission maps from radiative transfer calculations are convolved with a circular Gaussian kernel (i.e., beam) to mimic the angular resolution in interferometric observations.

\subsubsection{Parameter Space}

We explore a parameter space of three factors: dust mass, dust distribution, and the full width at half maximum (FWHM) of the interferometric beam. They are summarized in Table \ref{tab:rt_param}.

\paragraph{Dust mass} This parameter controls the total mass of each dust species within the numerical domain of calculation. Among the three dust species in our radiative transfer calculations, the micron-sized one is only used for a realistic temperature profile in the disk and is not expected to be bright at 1.3 mm. Thus, its total mass is fixed to $M_{\rm 1\mu m} = 10^{-6}M_\star$ for all models\footnote{In Appendix \ref{app:temp}, we show that other than changing the overall brightness, $M_{\rm 1\mu m}$ has negligible impacts on the dust substructure of interest. Therefore, it is not included in the parameter survey. For a similar reason, $T_\star$ is also not included in the parameter survey.}. For the other two dust species, we explore combinations of (1) $M_{\rm 100\mu m} = 10^{-4}M_\star$ and $ M_{\rm 1cm} = 10^{-3}M_\star$; (2) $M_{\rm 100\mu m} = M_{\rm 1cm} = 10^{-3}M_\star$; and (3) $M_{\rm 100\mu m} = 10^{-3}M_\star$ and $ M_{\rm 1cm} = 10^{-4}M_\star$, aligning with the models described in Section \ref{sec:dust_init}.

\paragraph{Dust distribution} The dust emission profile depends on the spatial distribution of dust. In this work, this factor is studied by changing the planet's mass from Saturn-like ($3\times10^{-4}M_\star$) to Jupiter-like ($10^{-3}M_\star$) in our hydrodynamic simulations, which end up with drastically different density profiles. 

\paragraph{Beam size} We perform convolution to the flux density profile calculated by \textsc{radmc3d} to mimic interferometric observations. The profiles are convolved with a circular Gaussian beam. The explored beam sizes (FWHM) are 25, 50, and 100 mas, corresponding to linear sizes of 3.5, 7, and 14 au at 140 pc. 

\begin{table*}[t]
  \setlength\tabcolsep{0pt}
  \centering
  \caption{Parameter Space for RT Calculations}\label{tab:rt_param}\vspace{-8pt}
  \begin{tabular}{C{9}C{1}C{1}C{1}C{1}C{1}C{1}C{1}C{1}C{1}C{1}C{1}C{1}C{1}C{1}C{1}C{1}C{1}C{1}C{1}C{1}C{1}C{1}C{1}C{1}C{1}C{1}C{1}C{1}C{1}C{1}}
  %   \hline
  %   \multicolumn{1 }{C{9 }|}{\multirow{3}{*}{Star}}
  % & \multicolumn{15}{C{15}|}{T Tauri (PDS 70)}
  % & \multicolumn{15}{C{15} }{HAeBe (AB Aur)} \\
  %   \multicolumn{1 }{C{9 }|}{\hfill}
  % & \multicolumn{15}{C{15}|}{$M_\star = 0.76 M_\odot$}
  % & \multicolumn{15}{C{15} }{$M_\star = 2.36 M_\odot$} \\
  %   \multicolumn{1 }{C{9 }|}{\hfill}
  % & \multicolumn{15}{C{15}|}{$T_\star = 3972{\rm K}$}
  % & \multicolumn{15}{C{15} }{$T_\star = 9770{\rm K}$} \\
    \hline
    \multicolumn{1 }{C{9 }|}{\multirow{2}{*}{Planet}}
  & \multicolumn{15}{C{15}|}{Saturn-like}
  & \multicolumn{15}{C{15} }{Jupiter-like} \\
    \multicolumn{1 }{C{9 }|}{\hfill}
  & \multicolumn{15}{C{15}|}{$3\times10^{-4}M_\star$}
  & \multicolumn{15}{C{15} }{$1\times10^{-3}M_\star$} \\
    \hline
    \multicolumn{1 }{C{9 }|}{\multirow{2}{*}{Dust Mass}}
  & \multicolumn{10}{C{10}|}{$M_{\rm 100\mu m} = 10^{-4}M_\star$}
  & \multicolumn{10}{C{10}|}{$M_{\rm 100\mu m} = 10^{-3}M_\star$}
  & \multicolumn{10}{C{10} }{$M_{\rm 100\mu m} = 10^{-3}M_\star$} \\
    \multicolumn{1 }{C{9 }|}{\hfill}
  & \multicolumn{10}{C{10}|}{$M_{\rm 1cm} = 10^{-3}M_\star$}
  & \multicolumn{10}{C{10}|}{$M_{\rm 1cm} = 10^{-3}M_\star$}
  & \multicolumn{10}{C{10} }{$M_{\rm 1cm} = 10^{-4}M_\star$} \\
    \hline
    \multicolumn{1 }{C{9 }|}{\multirow{2}{*}{Beam Size}}
  & \multicolumn{10}{C{10}|}{25 mas}
  & \multicolumn{10}{C{10}|}{50 mas}
  & \multicolumn{10}{C{10} }{100 mas} \\
    \multicolumn{1 }{C{9 }|}{\hfill}
  & \multicolumn{10}{C{10}|}{(3.5 au)}
  & \multicolumn{10}{C{10}|}{(7 au)}
  & \multicolumn{10}{C{10} }{(14 au)} \\
    \hline
  \end{tabular}
  % \justify \vspace{-8pt}
  % \tablecomments{}
\end{table*}
\section{Results} \label{sec:result}

\subsection{The Radial Migration of Dust Rings}

\begin{figure*}[t]
\centering
\includegraphics[width = \textwidth]{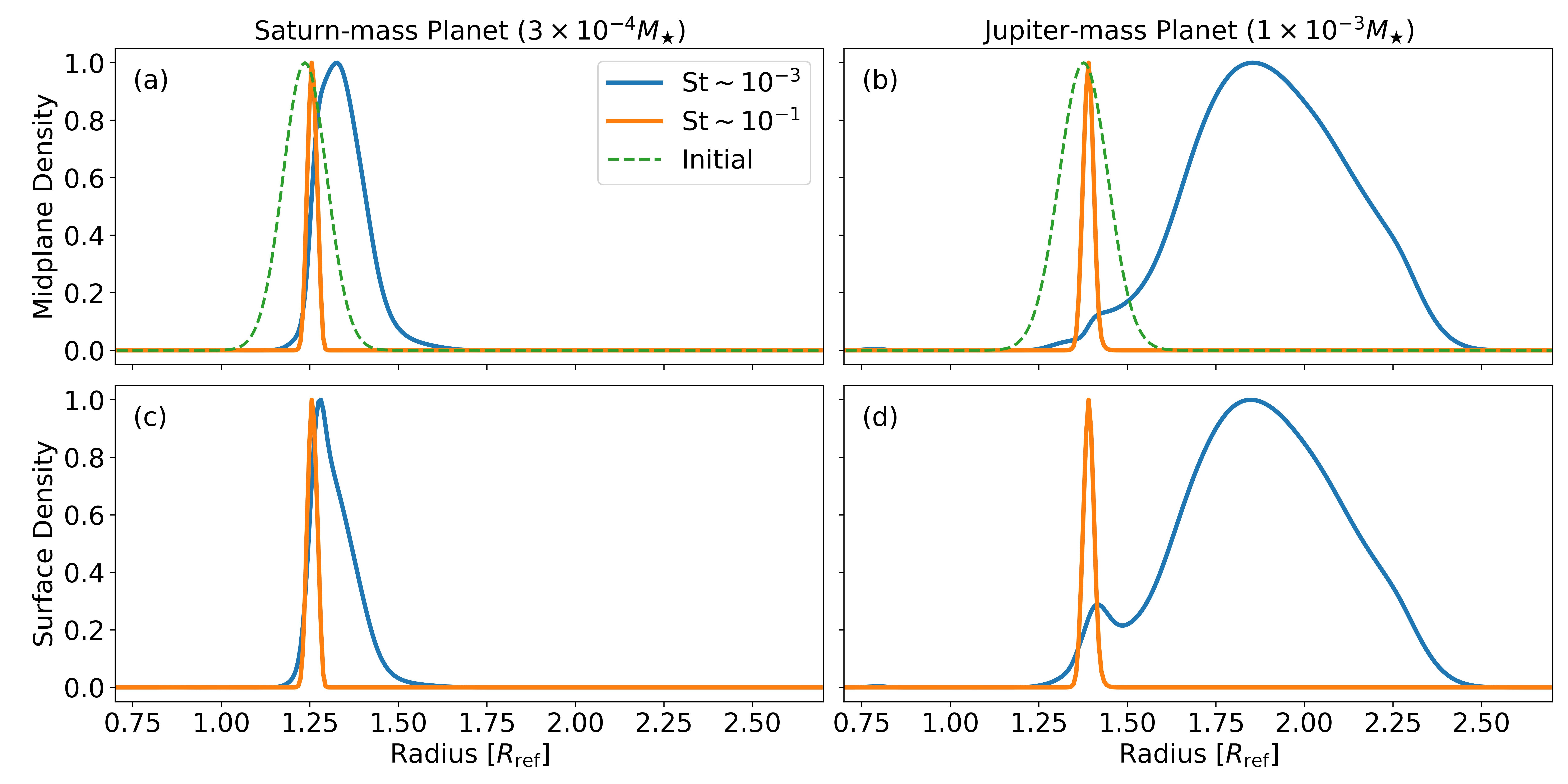}
\figcaption{
The normalized radial profiles of the azimuthally averaged dust density from the models with $\epsilon_0({\rm St_{ref} = 10^{-3}}) = 0.1$ and $\epsilon_0({\rm St_{ref} = 10^{-1}}) = 1$. The top panels are volumetric density on the midplane, and the bottom panels are the vertically integrated surface density. The left panels are from the Saturn-mass planet model, and the right panels are from the Jupiter-mass planet model. The initial dust profiles (same for both dust species) are plotted in the top panels. The planet is at $R = 1\,R_{\rm ref}$. 
\label{fig:dens}}
\end{figure*}

Figure \ref{fig:dens} summarizes the radial profile of dust density from our hydrodynamic simulations. In agreement with \citealt{bi_gap-opening_2023}, the dust ring composed of small dust grains is more radially extended than that of large grains in both models. Moreover, we notice that the dust ring of small grains is at a larger radius, even though the two dust species share the same initialization. 

The comparison between the density profiles and their initialization in panels (a) and (b) in Figure \ref{fig:dens} shows that the large grains are still concentrated in the local pressure maximum, whereas the small grains have migrated outward. The radial migration of the dust rings of smaller grains is only tentatively seen in the Saturn-mass case but is much more prominent in the Jupiter-mass planet case. This explains why such a feature was not confidently reported in our previous works, which focused on the gap-opening process by planets of $\lesssim$ the thermal mass\footnote{The thermal mass refers to the mass of a planet whose Hill sphere equals the gas scale height. In our study, $M_{\rm th} = 3(H_{\rm g}/R)^3M_\star \sim 4\times 10^{-4}M_\star$.} (e.g., Saturn-mass).

\begin{figure*}[t]
\centering
\includegraphics[width = \textwidth]{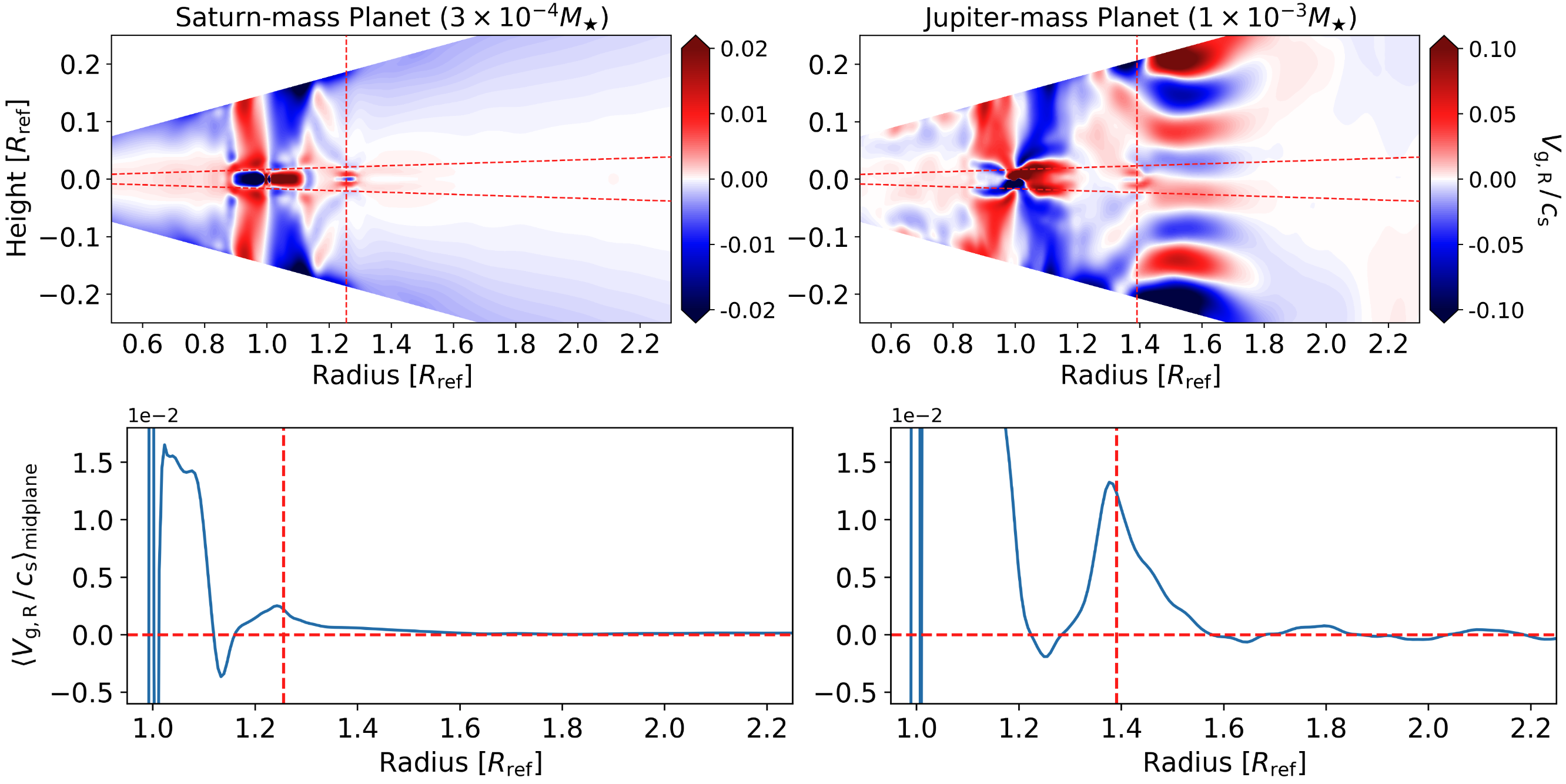}
\figcaption{The azimuthally averaged gas cylindrical radial velocity from the models with $\epsilon_0({\rm St_{ref} = 10^{-3}}) = 0.1$ and $\epsilon_0({\rm St_{ref} = 10^{-1}}) = 1$. All values are weighed by the gas density and are normalized to the local sound speed. The left panels are from the Saturn-mass planet model, and the right panels are from the Jupiter-mass planet model. The bottom panels are vertically averaged over a range of $\pm$\sfrac{1}{3}$H_{\rm g}$ around the midplane, which is marked in the top panels. The vertical dash line marks the radial location of the local pressure maximum in each model. The planet is at $R = 1\,R_{\rm ref}$.
\label{fig:vel}}
\end{figure*}

We find the origin of the outward radial migration of dust rings to be the gas flows driven by the planet, which are illustrated in Figure \ref{fig:vel}. The top panels, demonstrating the radial velocity of gas, show an outward flow at the local pressure maximum close to the midplane. As a result, the well-coupled, small dust grains are carried outward by the gas flows. In contrast, the loosely coupled, large dust grains remain concentrated at the local pressure maximum. This particular flow pattern is more recognizable in the $\lesssim$ thermal mass (i.e., Saturn-mass) case and becomes less organized in the > thermal mass (i.e., Jupiter-mass) case. Nevertheless, the outward flows can still be seen in the density-weighted vertically averaged plots over a thin layer on the midplane, shown in the bottom panels. It also shows that the Jupiter-mass planet drives a stronger flow in the midplane, which explains why the dust ring is at a larger radius than the Saturn-mass case. 

Although the outward gas flows on the midplane of the outer gap edge span a sizeable radial range of a few $H_{\rm g}$, we do not expect them to change the overall gas density profile (i.e., the surface density profile) as they do on the dust. This is because, in our 3D models, the outward gas flows on the midplane are balanced by the inward gas flows above the midplane, as illustrated in the top panels of Figure \ref{fig:vel}. Meanwhile, we do not expect a similar balancing effect on the dust. Once carried out of the gap edge, where the puff-up happens, the small dust grains will settle back to the midplane, and the replenishing gas flows high above will not be able to bring all of the dust back.

Returning to Figure \ref{fig:dens}, we notice the difference between the midplane density and surface density profiles, namely the excess of small dust grains close to the local pressure maximum. Those dust grains are stored above the disk midplane due to the puff-up at the outer gap edge, agreeing with \citealt{bi_puffed-up_2021}. 

\subsection{The Formation of Shoulders}

\begin{figure*}[t]
\centering
\includegraphics[width = \textwidth]{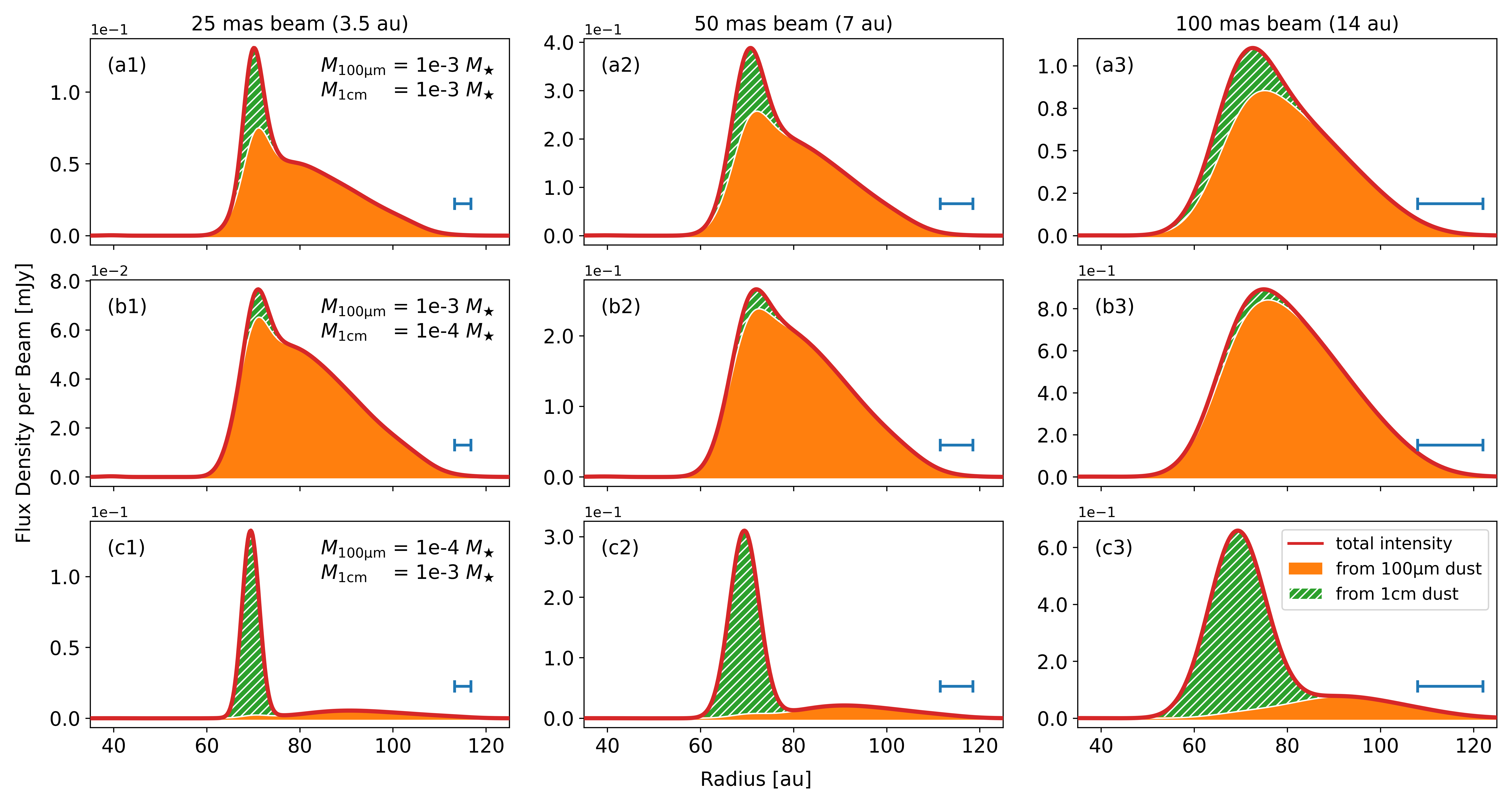}
\figcaption{The azimuthally averaged radial profiles of the dust continuum emission (flux density per beam) at 1.3 mm wavelength from radiative transfer calculations. All panels are from models with a Jupiter-mass planet. The shaded regions below each curve demonstrate the contribution of different dust species to the total intensity, with a legend in the bottom right panel. Panels on the same row share the same grain size distribution labeled on the top right. Panels in the same column are processed with the same beam size labeled on the top. The beam size is also labeled on the bottom right of each panel for reference. The planet is at 50 au.
\label{fig:jup}}
\end{figure*}

\begin{figure*}[t]
\centering
\includegraphics[width = \textwidth]{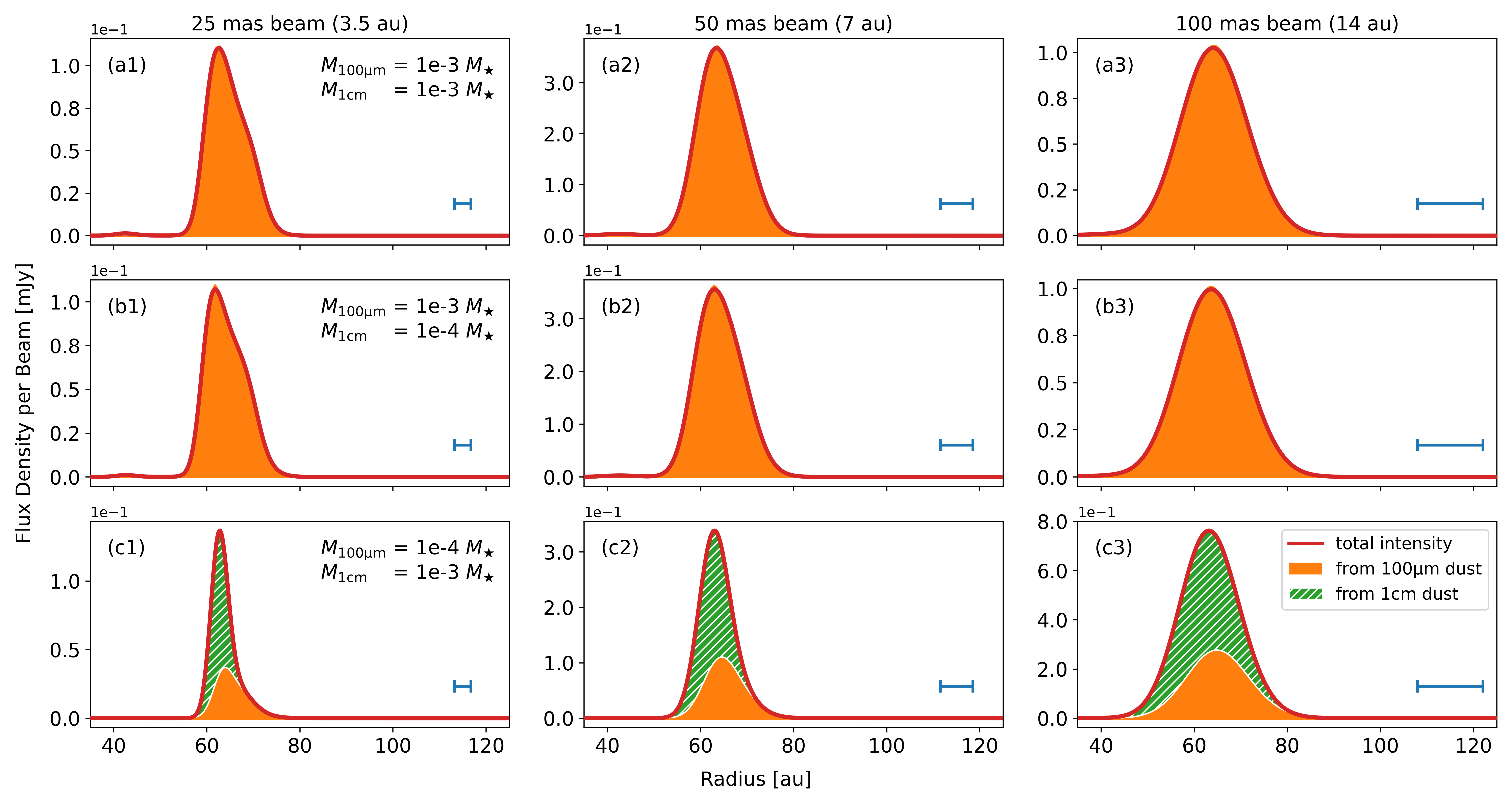}
\figcaption{Similar to Figure \ref{fig:jup}, but for models with a Satrun-mass planet.
\label{fig:sat}}
\end{figure*}

The radial migration of the dust ring leads to a change in the dust emission profile. Here, we let the physical grain size of the small and large grains be 100 $\mu$m and 1 cm, respectively and examine the dust continuum emission at 1.3 mm wavelength from our hydrodynamic models (see Table \ref{tab:rt_param} for the parameter space and Section \ref{sec:radmc3d} for descriptions and justifications). The results are summarized in Figure \ref{fig:jup} and \ref{fig:sat}. 

\subsubsection{The Jupiter-mass planet case} \label{sec:jup}

In the Jupiter-mass planet case (Figure \ref{fig:jup}), we find that the separated concentrations of small and large dust grains form ring-shoulder pairs or even separated rings in the emission profile (however, we still call them ring-shoulder pairs in this study for simplicity). The comparison among rows in Figure \ref{fig:jup} shows that the mass ratio between the two dust species is a major factor affecting the flux density ratio between the shoulder and the ring.

Besides that, we find the beam size (i.e., resolution) to be another factor. The comparison among columns in Figure \ref{fig:jup} shows that the ring-shoulder flux density ratio decreases when the beam size increases. The most prominent contrast can be found in row (c). This is because the concentration of large grains is an intrinsically sharper substructure compared with that of small grains. Therefore, the more emission from large grains, the more reduction of the peak value when convolved with large beams, which is the case for row (c). The reason larger grains have sharper concentrations was discussed in \citealt{bi_gap-opening_2023}, which is that small dust grains are more susceptible to the effective diffusion driven by the planet-disk interactions and thus have a wider radial distribution.

\subsubsection{The Saturn-mass planet case}

In the Saturn-mass planet case (Figure \ref{fig:sat}), we do not see any separated rings, and the shoulder can only be tentatively seen with a small beam size (panels (a1) and (b1)). This is because the gas flows driven by the Saturn-mass planet are not strong enough to move small dust grains out of the local pressure maximum. 

\begin{figure}[t]
\centering
\vspace{10pt}
\includegraphics[width = 0.47\textwidth]{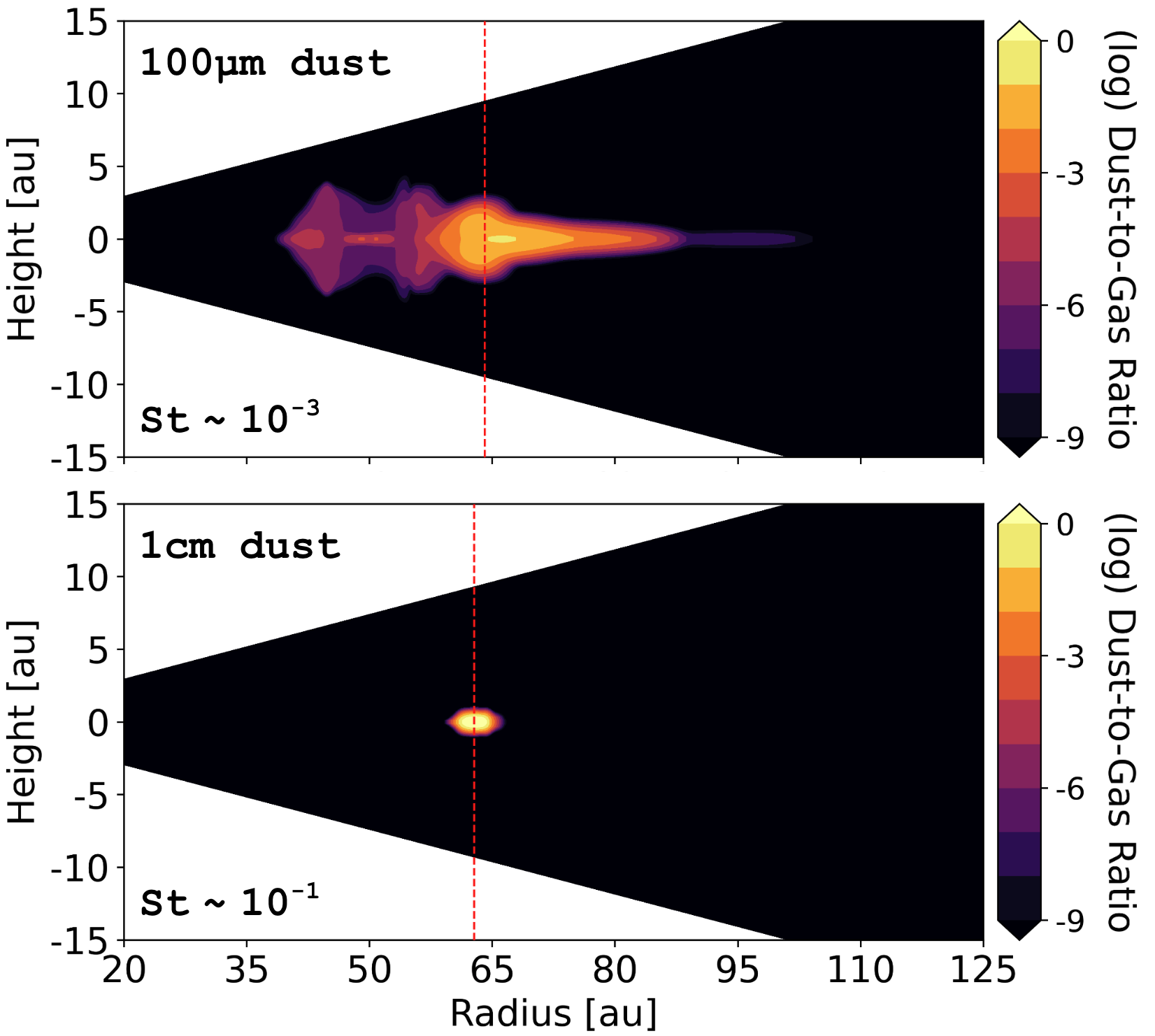}
\figcaption{The azimuthally averaged dust-to-gas density ratio in the hydrodynamic simulation of $\epsilon_0({\rm St_{ref} = 10^{-3}}) = 0.1$ and $\epsilon_0({\rm St_{ref} = 10^{-1}}) = 1$ with a Saturn-mass planet. The top panel is for the 100 $\mu$m (${\rm St}\sim 10^{-3}$) dust, and the bottom panel is for the 1 cm (${\rm St}\sim 10^{-1}$) dust. The vertical dashed line marks the radius of the maximum dust surface density of the corresponding dust species. The planet is at 50 au.
\label{fig:dgratio}}
\end{figure}

Besides, the first two rows show that the emission is dominated by the sub-mm dust grains unless the cm-sized ones are much more abundant (e.g., the third row). This is because the small dust grains trapped at the local pressure maximum are subject to the strong puff-up effect discussed in \citealt{bi_puffed-up_2021} and illustrated in Figure \ref{fig:dgratio}. Therefore, they are carried upward by the gas flows driven by the planet and become warmer and much brighter than the large grains settled on the midplane.
\section{Discussion} \label{sec:discuss}

\subsection{The Outer Shoulders in Disk Surveys}

In \citealt{jennings_super-resolution_2022} and \citealt{jennings_superresolution_2022}, shoulders attached to the outer part of dust rings, as well as adjacent rings with shallow gaps in between, are found in multiple disks such as HD 163296 and AS 209. Our 3D planet-disk models featuring multiple dust species capture these features. Therefore, we argue that these outer shoulders can be natural outcomes of the ongoing planet-disk interactions with $\gtrsim$ thermal-mass planet(s) there. Besides, we show that the structure of the ring-shoulder pair is sensitive to the grain size distribution, the planet's mass, and the beam size in observations. The last factor is known in observations; therefore, we argue that our study opens a new avenue to probe the planet's mass and the grain size distribution in disks with shoulders found. Further observations, especially multi-wavelength dust continuum observations, are needed to constrain the grain size distribution in the disk and test our theory. 

\subsection{The Inner Shoulder in PDS 70}

\begin{figure}[t]
\centering
\includegraphics[width = 0.47\textwidth]{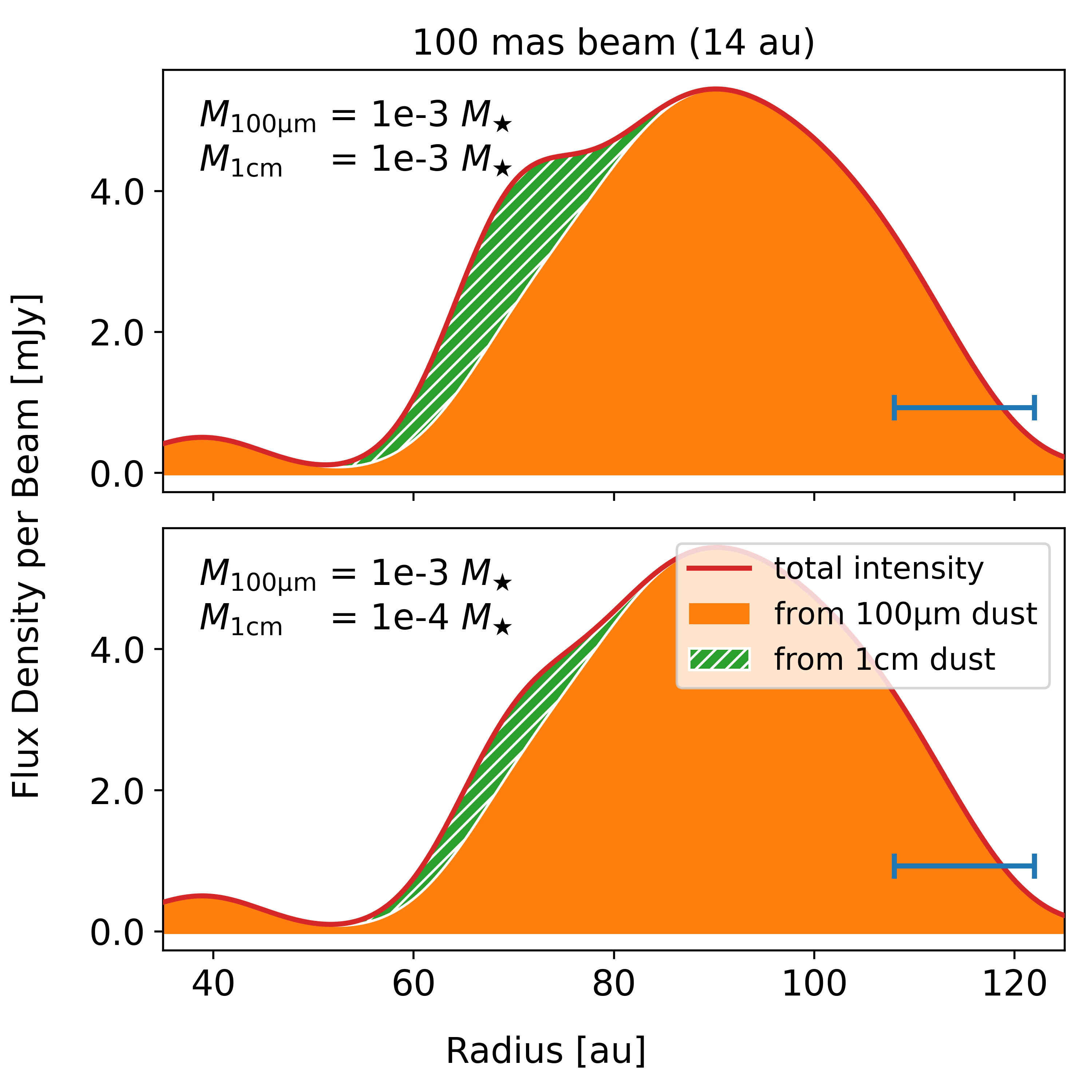}
\figcaption{Similar to Figure \ref{fig:jup}, but dust feedback on the gas is neglected in the hydrodynamic simulations. The top and bottom panels correspond to panels (a3) and (b3) of Figure \ref{fig:jup}.
\label{fig:nofb}}
\end{figure}

In \citealt{benisty_circumplanetary_2021}, an inner shoulder is reported at the outer edge of the gap opened by planets. However, our simulations did not capture such a feature. One of the possible reasons is that the properties of the PDS 70 planet-disk system reside beyond our parameter space. Another possible reason is the effectiveness of dust feedback on the gas. In \citealt{bi_gap-opening_2023}, we showed that dust feedback plays an important role in dust dynamics when effective. In this work, if we neglect dust feedback by assuming it is ineffective, we can find inner shoulders showing up in some of our simulations (see Figure \ref{fig:nofb}). However, this poses additional constraints, indicating that the disk is typically massive or dust is not abundant. Overall, whether the formation of the inner shoulder in the PDS 70 disk is associated with the ongoing planet-disk interactions is still to be investigated by future case studies on this target.

\subsection{Caveats and Outlooks}

In this study, we consider only two representative dust grain sizes. This prescription of grain size distribution is sufficient for our proof-of-concept study, which shows that planet-disk interactions can move small dust grains out of the local pressure maximum and change the emission profile correspondingly. However, two dust grain sizes separated by two orders of magnitude may still be too coarse to thoroughly investigate the effect of planet-disk interactions on dust emission because both dust dynamics in the disk and the dust emissivity depend on the grain size. Therefore, more dust species are needed in future studies to understand this problem better. In the meantime, increasing dust species means increasing the computational expense. We thus highlight the importance of massively paralleled methods performed by GPU and alternative prescriptions of the grain size distribution, such as tracking the one characteristic/dominant grain size at each location (e.g., \citealt{ormel_atmospheric_2014}). 

Another potential caveat of the study is the non-flared disk geometry adopted when radiative transfer processes are involved. The non-flared geometry in hydrodynamic simulations is convenient because the vertical domain stops at a particular gas scale height. It provides a constant vertical resolution per scale height in the spherical coordinates and, more importantly, avoids calculating the low-density region in the inner disk above 4--5 scale heights, thereby greatly accelerating the calculations. However, the non-flared disk is self-shadowed when radiative transfer processes are considered because the central star cannot directly illuminate its surface layer. Therefore, we may have underestimated the dust temperature, especially on the midplane far from the planet-opened gap, which may further affect the dust thermal emission profile. In addition, to reach a reasonable flux density level of the ring-shoulder pair\footnote{Here, we refer to the flux density level of the ring-shoulder pair in the ALMA observation of PDS 70 published in \citealt{benisty_circumplanetary_2021}, which is on the order of 0.1 mJy per roughly 30 mas by 30 mas beam.}, a relatively high dust mass is needed in this study due to the underestimation of temperature. Therefore, we caution that the disk geometry needs to be taken better care of in future works, especially when aiming to constrain any disk properties (e.g., dust mass) from the flux density level in observations.

\section{Conclusion} \label{sec:conclusion}

In this work, we combine 3D hydrodynamic simulations with radiative transfer calculations to seek explanations for the shoulders of dust rings found previously in mm-wavelength disk observations. Our main findings are:

\begin{itemize}

\item Dust shoulders attached to the outer edge of the rings may be outcomes of 3D planet-disk interactions with massive, gap-opening planets. Our study may explain the outer shoulders frequently seen in super-resolution analyses of ALMA dust rings, whose origins remained unclear. Therefore, we highlight the importance of 3D analyses when studying dust dynamics under planet-disk interactions.

\item The key driver for forming the ring-shoulder pair is the filtration effect of dust due to planet-driven outward gas flows at the local pressure maximum. Sub-mm-sized dust grains are carried outward by the gas flows, whereas cm-sized ones are not, resulting in separated concentrations of dust that give rise to multiple peaks in the dust continuum emission profile. This mechanism is prominent when the planet is > thermal mass and becomes less effective when the planet is $\lesssim$ thermal mass.

\item Our parameter survey suggests the flux density ratio between the shoulder and the ring is sensitive to the grain size distribution and the planet's mass. The ratio can also be affected by the angular resolution of the observation.

\item While our fiducial models do not recover the inner shoulder found in the planet-hosting disk PDS 70, we provide a scenario where the dust feedback on the gas is neglected, in which an inner shoulder may appear. We thus suggest that more detailed studies are needed to better understand the dynamical processes in this disk.

\end{itemize}

We thank Ruobing Dong, Kees Dullemond, Jeffrey Fung, Thomas Rometsch, Richard Teague, and Yanqin Wu for the helpful discussions and the anonymous referee for the insightful comments that greatly improved this study. Simulations were carried out on the KAWAS cluster hosted by ASIAA. MKL is supported by the National Science and Technology Council (grants 111-2112-M-001-062-, 112-2112-M-001-064-, 111-2124-M-002-013-, 112-2124-M-002-003-) and an Academia Sinica Career Development Award (AS-CDA110-M06).

\clearpage
\appendix

\section{The Time Evolution of Small Dust Grains} \label{app:time}

\begin{figure*}[h]
\centering
\includegraphics[width = \textwidth]{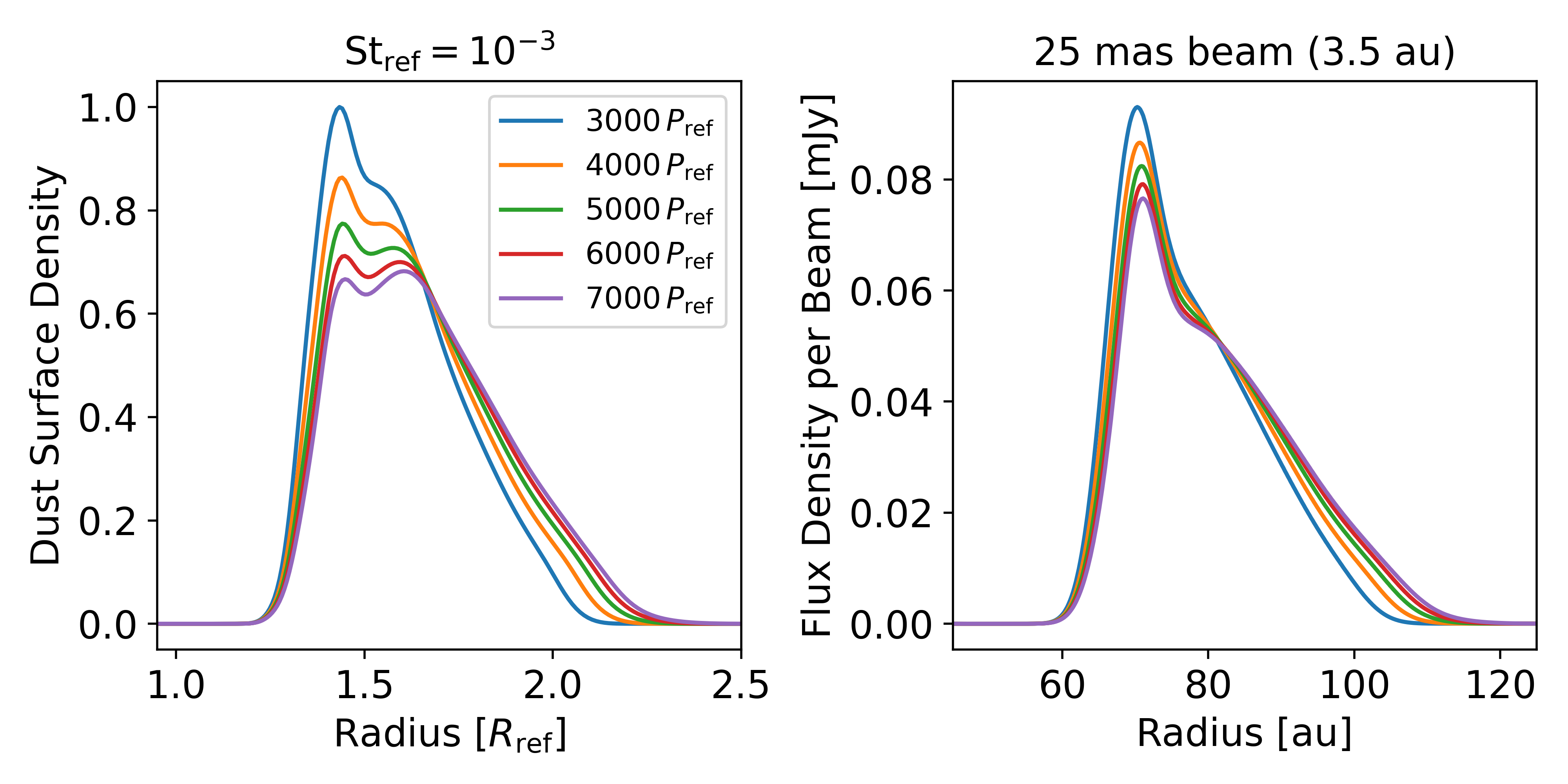}
\figcaption{The time evolution of the azimuthally averaged surface density profile of the small dust grains from hydrodynamic simulations (left) and the total dust continuum emission from radiative transfer calculations (right). The hydrodynamic simulations adopt a Jupiter-mass planet, $\epsilon_0({\rm St_{ref} = 10^{-3}}) = 1$, and $\epsilon_0({\rm St_{ref} = 10^{-1}}) = 0.1$. The radiative transfer calculations adopt a Jupiter-mass planet, $M_{\rm 100\mu m} = 10^{-3}M_\star$, and $M_{\rm 1cm} = 10^{-4}M_\star$. 
\label{fig:Time}}
\end{figure*}

The left panel of Figure \ref{fig:Time} shows the time evolution of the azimuthally averaged surface density profile of small dust grains from the hydrodynamic simulation with a Jupiter-mass planet, $\epsilon_0({\rm St_{ref} = 10^{-3}}) = 1$, and $\epsilon_0({\rm St_{ref} = 10^{-1}}) = 0.1$. The corresponding results in radiative transfer calculations are shown in the right panel. Despite the small grains not reaching a static state at 7000$P_{\rm ref}$, the dust emission is not strongly affected. We, therefore, suggest that the non-static states of dust in our hydrodynamic simulations do not affect our interpretations significantly.

\section{The effect of the micron-sized dust in the parameter survey} \label{app:temp}

\begin{figure*}[h]
\centering
\includegraphics[width = \textwidth]{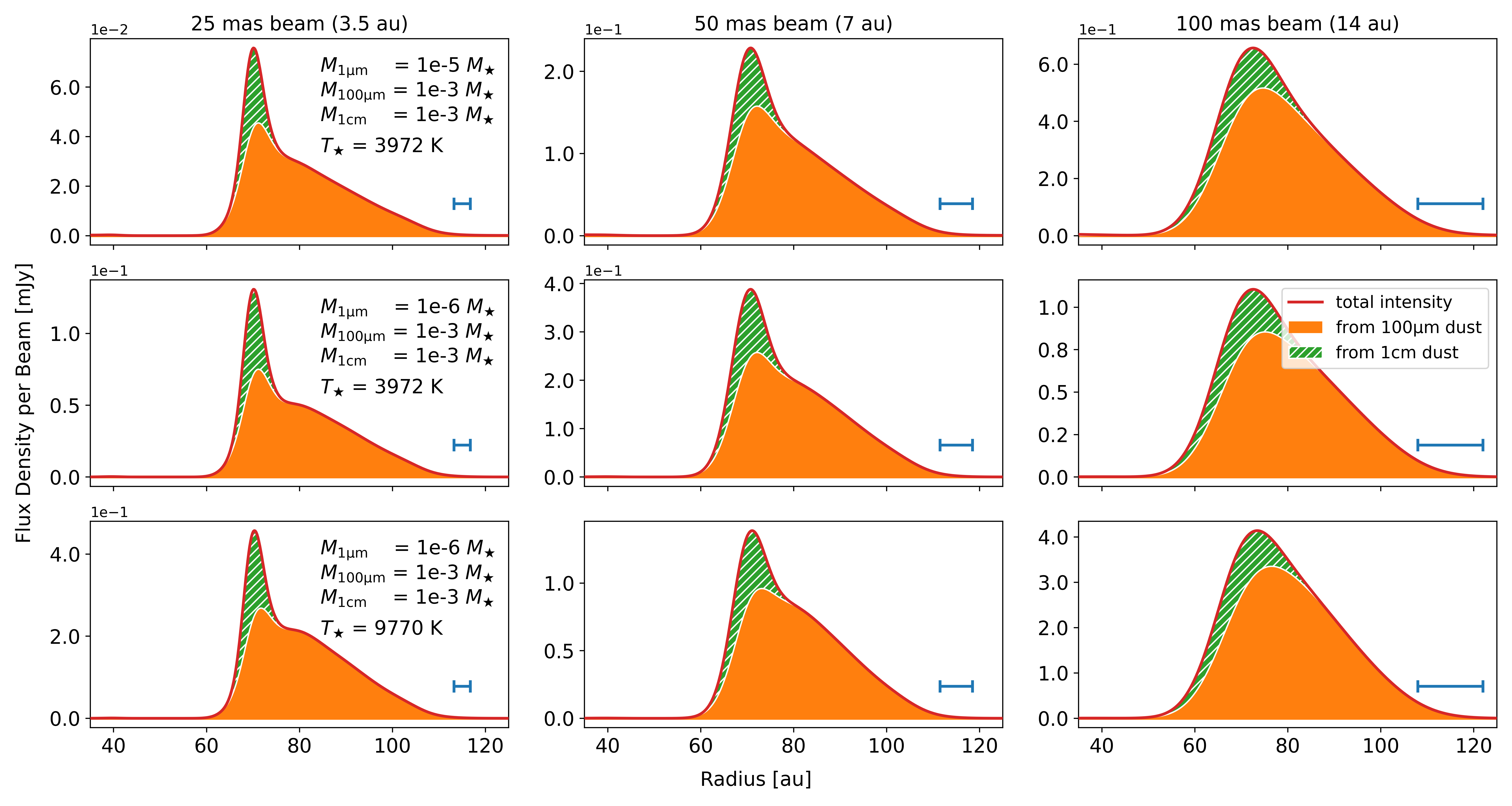}
\figcaption{The azimuthally averaged radial profiles of the dust continuum emission (flux density per beam) at 1.3 mm wavelength from radiative transfer calculations. All panels are from models of $M_{\rm 100\mu m} = M_{\rm 1cm} = 10^{-3}M_\star$ with a Jupiter-mass planet. The shaded regions below each curve demonstrate the contribution of different dust species to the total intensity. The middle panels are from models of $M_{\rm 1\mu m} = 10^{-6}M_\star$, same as the top panels in Figure \ref{fig:jup}. Compared with the middle panels, the top panels have a higher $M_{\rm 1\mu m}$ and the bottom panels have a higher $T_\star$. Panels in the same column are processed with the same beam size labeled on the top. The beam size is also labeled on the bottom right of each panel for reference. 
\label{fig:temp}}
\end{figure*}

Figure \ref{fig:temp} shows the radial profiles of dust emission when the mass of the micron-sized dust or the blackbody temperature of the star is changed. The comparison between the first and the second row shows that changing $M_{\rm 1\mu m}$ does not significantly affect the structure of the ring-shoulder pair, except for changing the overall intensity. The comparison between the second and the third row shows that changing $T_\star$ also does not affect the overall structure. Therefore, we do not include $M_{\rm 1\mu m}$ and $T_\star$ in our parameter survey in this work and only use them to specify a disk temperature profile.
\bibliography{jiaqing}

\begin{thebibliography}{}
\expandafter\ifx\csname natexlab\endcsname\relax\def\natexlab#1{#1}\fi

\bibitem[{Benisty {et~al.}(2021)Benisty, Bae, Facchini, Keppler, Teague, Isella, Kurtovic, Pérez, Sierra, Andrews, Carpenter, Czekala, Dominik, Henning, Menard, Pinilla, \& Zurlo}]{benisty_circumplanetary_2021}
Benisty, M., Bae, J., Facchini, S., {et~al.} 2021, The Astrophysical Journal, 916, L2

\bibitem[{Benítez-Llambay {et~al.}(2019)Benítez-Llambay, Krapp, \& Pessah}]{benitez-llambay_asymptotically_2019}
Benítez-Llambay, P., Krapp, L., \& Pessah, M.~E. 2019, The Astrophysical Journal Supplement Series, 241, 25

\bibitem[{Benítez-Llambay \& Masset(2016)}]{benitez-llambay_fargo3d_2016}
Benítez-Llambay, P., \& Masset, F.~S. 2016, The Astrophysical Journal Supplement Series, 223, 11

\bibitem[{Bi {et~al.}(2021)Bi, Lin, \& Dong}]{bi_puffed-up_2021}
Bi, J., Lin, M.-K., \& Dong, R. 2021, The Astrophysical Journal, 912, 107

\bibitem[{Bi {et~al.}(2023)Bi, Lin, \& Dong}]{bi_gap-opening_2023}
---. 2023, The Astrophysical Journal, 942, 80

\bibitem[{Birnstiel(2023)}]{birnstiel_dust_2023}
Birnstiel, T. 2023, Dust growth and evolution in protoplanetary disks, arXiv:2312.13287

\bibitem[{Birnstiel {et~al.}(2018)Birnstiel, Dullemond, Zhu, Andrews, Bai, Wilner, Carpenter, Huang, Isella, Benisty, Pérez, \& Zhang}]{birnstiel_disk_2018}
Birnstiel, T., Dullemond, C.~P., Zhu, Z., {et~al.} 2018, The Astrophysical Journal, 869, L45

\bibitem[{Dominik {et~al.}(2021)Dominik, Min, \& Tazaki}]{dominik_optool_2021}
Dominik, C., Min, M., \& Tazaki, R. 2021, Astrophysics Source Code Library, ascl:2104.010

\bibitem[{Drążkowska {et~al.}(2023)Drążkowska, Bitsch, Lambrechts, Mulders, Harsono, Vazan, Liu, Ormel, Kretke, \& Morbidelli}]{drazkowska_planet_2023}
Drążkowska, J., Bitsch, B., Lambrechts, M., {et~al.} 2023, Protostars and Planets VII, 534, 717

\bibitem[{Dullemond {et~al.}(2012)Dullemond, Juhasz, Pohl, Sereshti, Shetty, Peters, Commercon, \& Flock}]{dullemond_radmc-3d_2012}
Dullemond, C.~P., Juhasz, A., Pohl, A., {et~al.} 2012, Astrophysics Source Code Library, ascl:1202.015

\bibitem[{Dullemond \& Penzlin(2018)}]{dullemond_dust-driven_2018}
Dullemond, C.~P., \& Penzlin, A. B.~T. 2018, Astronomy and Astrophysics, 609, A50

\bibitem[{Fung \& Chiang(2016)}]{fung_gap_2016}
Fung, J., \& Chiang, E. 2016, The Astrophysical Journal, 832, 105

\bibitem[{Hayashi(1981)}]{hayashi_structure_1981}
Hayashi, C. 1981, Progress of Theoretical Physics Supplement, 70, 35

\bibitem[{Henyey \& Greenstein(1941)}]{henyey_diffuse_1941}
Henyey, L.~G., \& Greenstein, J.~L. 1941, The Astrophysical Journal, 93, 70

\bibitem[{Huang {et~al.}(2018)Huang, Andrews, Dullemond, Isella, Pérez, Guzmán, Öberg, Zhu, Zhang, Bai, Benisty, Birnstiel, Carpenter, Hughes, Ricci, Weaver, \& Wilner}]{huang_disk_2018-1}
Huang, J., Andrews, S.~M., Dullemond, C.~P., {et~al.} 2018, The Astrophysical Journal, 869, L42

\bibitem[{Huang {et~al.}(2020)Huang, Andrews, Dullemond, Öberg, Qi, Zhu, Birnstiel, Carpenter, Isella, Macías, McClure, Pérez, Teague, Wilner, \& Zhang}]{huang_multifrequency_2020}
---. 2020, The Astrophysical Journal, 891, 48

\bibitem[{Jennings {et~al.}(2022{\natexlab{a}})Jennings, Booth, Tazzari, Clarke, \& Rosotti}]{jennings_super-resolution_2022}
Jennings, J., Booth, R.~A., Tazzari, M., Clarke, C.~J., \& Rosotti, G.~P. 2022{\natexlab{a}}, Monthly Notices of the Royal Astronomical Society, 509, 2780

\bibitem[{Jennings {et~al.}(2020)Jennings, Booth, Tazzari, Rosotti, \& Clarke}]{jennings_frankenstein_2020}
Jennings, J., Booth, R.~A., Tazzari, M., Rosotti, G.~P., \& Clarke, C.~J. 2020, Monthly Notices of the Royal Astronomical Society, 495, 3209

\bibitem[{Jennings {et~al.}(2022{\natexlab{b}})Jennings, Tazzari, Clarke, Booth, \& Rosotti}]{jennings_superresolution_2022}
Jennings, J., Tazzari, M., Clarke, C.~J., Booth, R.~A., \& Rosotti, G.~P. 2022{\natexlab{b}}, Monthly Notices of the Royal Astronomical Society, 514, 6053

\bibitem[{Jiang \& Ormel(2021)}]{jiang_survival_2021}
Jiang, H., \& Ormel, C.~W. 2021, Monthly Notices of the Royal Astronomical Society, 505, 1162

\bibitem[{Johansen {et~al.}(2009)Johansen, Youdin, \& Klahr}]{johansen_zonal_2009}
Johansen, A., Youdin, A., \& Klahr, H. 2009, The Astrophysical Journal, 697, 1269

\bibitem[{Keppler {et~al.}(2018)Keppler, Benisty, Müller, Henning, van Boekel, Cantalloube, Ginski, van Holstein, Maire, Pohl, Samland, Avenhaus, Baudino, Boccaletti, de~Boer, Bonnefoy, Chauvin, Desidera, Langlois, Lazzoni, Marleau, Mordasini, Pawellek, Stolker, Vigan, Zurlo, Birnstiel, Brandner, Feldt, Flock, Girard, Gratton, Hagelberg, Isella, Janson, Juhasz, Kemmer, Kral, Lagrange, Launhardt, Matter, Ménard, Milli, Mollière, Olofsson, Pérez, Pinilla, Pinte, Quanz, Schmidt, Udry, Wahhaj, Williams, Buenzli, Cudel, Dominik, Galicher, Kasper, Lannier, Mesa, Mouillet, Peretti, Perrot, Salter, Sissa, Wildi, Abe, Antichi, Augereau, Baruffolo, Baudoz, Bazzon, Beuzit, Blanchard, Brems, Buey, De~Caprio, Carbillet, Carle, Cascone, Cheetham, Claudi, Costille, Delboulbé, Dohlen, Fantinel, Feautrier, Fusco, Giro, Gluck, Gry, Hubin, Hugot, Jaquet, Le~Mignant, Llored, Madec, Magnard, Martinez, Maurel, Meyer, Möller-Nilsson, Moulin, Mugnier, Origné, Pavlov, Perret, Petit, Pragt, Puget, Rabou, Ramos, Rigal, Rochat,
  Roelfsema, Rousset, Roux, Salasnich, Sauvage, Sevin, Soenke, Stadler, Suarez, Turatto, \& Weber}]{keppler_discovery_2018}
Keppler, M., Benisty, M., Müller, A., {et~al.} 2018, Astronomy and Astrophysics, 617, A44

\bibitem[{Keppler {et~al.}(2019)Keppler, Teague, Bae, Benisty, Henning, van Boekel, Chapillon, Pinilla, Williams, Bertrang, Facchini, Flock, Ginski, Juhasz, Klahr, Liu, Müller, Pérez, Pohl, Rosotti, Samland, \& Semenov}]{keppler_highly_2019}
Keppler, M., Teague, R., Bae, J., {et~al.} 2019, Astronomy and Astrophysics, 625, A118

\bibitem[{Lin \& Papaloizou(1993)}]{lin_tidal_1993}
Lin, D. N.~C., \& Papaloizou, J. C.~B. 1993, Protostars and Planets III, 749

\bibitem[{Long {et~al.}(2018)Long, Pinilla, Herczeg, Harsono, Dipierro, Pascucci, Hendler, Tazzari, Ragusa, Salyk, Edwards, Lodato, van~de Plas, Johnstone, Liu, Boehler, Cabrit, Manara, Menard, Mulders, Nisini, Fischer, Rigliaco, Banzatti, Avenhaus, \& Gully-Santiago}]{long_gaps_2018}
Long, F., Pinilla, P., Herczeg, G.~J., {et~al.} 2018, The Astrophysical Journal, 869, 17

\bibitem[{Müller {et~al.}(2018)Müller, Keppler, Henning, Samland, Chauvin, Beust, Maire, Molaverdikhani, van Boekel, Benisty, Boccaletti, Bonnefoy, Cantalloube, Charnay, Baudino, Gennaro, Long, Cheetham, Desidera, Feldt, Fusco, Girard, Gratton, Hagelberg, Janson, Lagrange, Langlois, Lazzoni, Ligi, Ménard, Mesa, Meyer, Mollière, Mordasini, Moulin, Pavlov, Pawellek, Quanz, Ramos, Rouan, Sissa, Stadler, Vigan, Wahhaj, Weber, \& Zurlo}]{muller_orbital_2018}
Müller, A., Keppler, M., Henning, T., {et~al.} 2018, Astronomy and Astrophysics, 617, L2

\bibitem[{Okuzumi {et~al.}(2016)Okuzumi, Momose, Sirono, Kobayashi, \& Tanaka}]{okuzumi_sintering-induced_2016}
Okuzumi, S., Momose, M., Sirono, S.-i., Kobayashi, H., \& Tanaka, H. 2016, The Astrophysical Journal, 821, 82

\bibitem[{Ormel(2014)}]{ormel_atmospheric_2014}
Ormel, C. 2014, The Astrophysical Journal, 789, L18

\bibitem[{Pecaut \& Mamajek(2016)}]{pecaut_star_2016}
Pecaut, M.~J., \& Mamajek, E.~E. 2016, Monthly Notices of the Royal Astronomical Society, 461, 794

\bibitem[{Regály \& Vorobyov(2017)}]{regaly_circumstellar_2017}
Regály, Z., \& Vorobyov, E. 2017, Astronomy and Astrophysics, 601, A24

\bibitem[{Shakura \& Sunyaev(1973)}]{shakura_black_1973}
Shakura, N.~I., \& Sunyaev, R.~A. 1973, Astronomy and Astrophysics, 24, 337

\bibitem[{Takahashi \& Inutsuka(2014)}]{takahashi_two-component_2014}
Takahashi, S.~Z., \& Inutsuka, S.-i. 2014, The Astrophysical Journal, 794, 55

\bibitem[{Takeuchi \& Artymowicz(2001)}]{takeuchi_dust_2001}
Takeuchi, T., \& Artymowicz, P. 2001, The Astrophysical Journal, 557, 990

\bibitem[{Zhang {et~al.}(2015)Zhang, Blake, \& Bergin}]{zhang_evidence_2015}
Zhang, K., Blake, G.~A., \& Bergin, E.~A. 2015, The Astrophysical Journal, 806, L7

\end{thebibliography}

%%%%%%%%%%%%%%%%%%%%%%%%%%%%%%%%%%%%%%%%%%%%%%%%%%%%%%%%%%%%%%%%%%%%%%%%%%%%%%%%
%%%%%%%%%%%%%%%%%%%%%%%%%%%%%%%%%%%%%%%%%%%%%%%%%%%%%%%%%%%%%%%%%%%%%%%%%%%%%%%%

\end{document}